\documentclass[aps,pra,twocolumn,superscriptaddress,longbibliography]{revtex4-1}

\usepackage{amsmath}
\usepackage{amssymb}
\usepackage{graphicx}
\usepackage[dvipsnames]{xcolor}
\usepackage[colorlinks,linkcolor=blue,citecolor=blue,urlcolor=blue]{hyperref}
\usepackage{stix}

\begin{document}

\renewcommand{\vec}[1]{\mathbf{#1}}
\newcommand{\gvec}[1]{\boldsymbol{#1}}
\newcommand{\iu}{\mathrm{i}}
\newcommand{\hc}{\hat{c}}
\newcommand{\hcd}{\hat{c}^\dagger}
\newcommand{\en}{\varepsilon}
\renewcommand{\pl}{\parallel}


\author{Michael Sch\"uler}
\email{schuelem@stanford.edu}
\affiliation{Stanford Institute for Materials and Energy Sciences (SIMES),
  SLAC National Accelerator Laboratory, Menlo Park, CA 94025, USA}
\author{Michael A.~Sentef}
\affiliation{Max Planck Institute for the Structure and Dynamics of
  Matter, Center for Free Electron Laser Science (CFEL), Luruper Chaussee 149, 22761 Hamburg, Germany}

\title{Theory of subcycle time-resolved photoemission: application to terahertz photodressing in graphene}

\begin{abstract}
  Motivated by recent experimental progress we revisit the theory of pump-probe time- and angle-resolved photoemission spectroscopy (trARPES), which is one of the most powerful techniques to trace transient pump-driven modifications of the electronic properties.
  The pump-induced dynamics can be described in different gauges for the light-matter interaction. Standard minimal coupling leads to the velocity gauge, defined by linear coupling to the vector potential. In the context of tight-binding (TB) models, the Peierls substitution is the commonly employed scheme for single-band models. Multi-orbital extensions -- including the coupling of the dipole moments to the electric field -- have been introduced and tested recently. In this work, we derive the theory of time-resolved photoemission within both gauges from the perspective of nonequilibrium Green's functions. This approach naturally incorporates the photoelectron continuum, which allows for a direct calculation of the observable photocurrent. Following this route we introduce gauge-invariant expressions for the time-resolved photoemission signal. The theory is applied to graphene pumped with short terahertz pulses, which we treat within a first-principles TB model. We investigate the gauge invariance and discuss typical effects observed in subcycle time-resolved photoemission. Our formalism is an ideal starting point for realistic trARPES simulations including scattering effects.
\end{abstract}

\maketitle

\section{Introduction}

Ultrafast material science is a thriving field of modern physics~\cite{basov_towards_2017,wang_theoretical_2018,de_la_torre_nonthermal_2021}, fueled by the impressive progress in creating ultrafast laser pulses and in time-resolved spectroscopy~\cite{smallwood_ultrafast_2016,rohde_time-resolved_2016,lv_angle-resolved_2019,lee_high_2020}. These advances enable to explore intriguing phenomena beyond linear response such as nonlinear Bloch oscillations~\cite{schubert_sub-cycle_2014,reimann_subcycle_2018} and light-engineered electronic properties~\cite{wang_observation_2013,mahmood_selective_2016,de_giovannini_monitoring_2016,hubener_creating_2017,reutzel_coherent_2020,schuler_how_2020-1}. In particular, light-inducing and probing on-demand topological states in graphene~\cite{oka_photovoltaic_2009,kitagawa_transport_2011,sentef_theory_2015} and other systems~\cite{wang_observation_2013,mahmood_selective_2016,claassen_all-optical_2016-1,hubener_creating_2017,topp_topological_2019,topp_all-optical_2018,rudner_band_2020,kandelaki_many-body_2018,oka_floquet_2019,kibis_floquet_2020,perfetto_floquet_2020}, has been a long-standing challenge. While the transient Hall effect~\cite{schuler_quench_2019,mciver_light-induced_2019} gives some hints on global properties, measuring the effective band structure upon periodic driving (Floquet bands) -- as enabled by pump-probe time-resolved angle-resolved photoemission spectroscopy (trARPES) -- yields a detailed microscopic picture. However, the interplay of decoherence~\cite{sato_floquet_2019,sato_microscopic_2019}, scattering effects~\cite{gierz_tracking_2015,schuler_how_2020-1,aeschlimann_survival_2021} and screening~\cite{keunecke_electromagnetic_2020} hamper the direct observation of Floquet states.

In search for more favorable regimes, using terahertz (THz) pulses has emerged as new direction. These low-frequency drives result in pronounced Floquet physics at lower field strength~\cite{mciver_light-induced_2019}, albeit scattering effects seem to play an elevated role~\cite{sato_microscopic_2019,aeschlimann_survival_2021}.
Floquet physics manifests by stroboscopic probing where the probe pulse averages over several optical cycles. Furthermore, using THz pump pulses (typical oscillation period of $\sim 20$~fs) -- combined with the femtosecond time resolution of typical trARPES setups~\cite{rohde_time-resolved_2016,lee_high_2020} -- allows for tracing subcycle information. Subcycle probing is complementary to the Floquet regime and provides insights into the build-up of photodressing effects.

To explore the transient photodressing as manifested in subcycle photoemission theoretically, few methods are available. Time-dependent density functional theory (TDDFT) provides a first-principle path, including a direct simulation of trARPES~\cite{de_giovannini_textitab_2012,de_giovannini_monitoring_2016}. However, the lack of electron-electron (beyond mean-field effects) or electron-phonon scattering underline the need for microscopic theories capable of including such effects. In this context, time-dependent nonequilibrium Green's functions (td-NEGF) approach~\cite{stefanucci_nonequilibrium_2013} has become one of the most powerful tools due to its natural connection to trARPES~\cite{freericks_theoretical_2009,sentef_examining_2013} and its flexibility to include various interaction effects. To the reduce the significant computational cost, constructing models in restricted band space is the standard route, often obtained from a tight-binding (TB) description of the relevant orbitals. The nonperturbative nature of photodressing requires incorporating light-matter interaction beyond linear response, which is difficult for empirically derived TB models. For instance, the straightforward Peierls substitution~\cite{peierls_zur_1933,ismail-beigi_coupling_2001} neglects local inter-orbital transitions, while introducing matrix elements of the light-matter coupling in the minimal coupling scheme directly from the TB model generally breaks gauge invariance~\cite{foreman_consequences_2002}. Note that violating gauge invariance can lead to qualitative artifacts like a spurious superradiant phase for spatially uniform fields~\cite{mazza_superradiant_2019}, which is impossible within a gauge-invariant description~\cite{andolina_cavity_2019,andolina_theory_2020}.

In contrast, TB models obtained by explicitly computing Wannier functions (WFs) from first-principle input contain the full orbital degrees of freedom, thus allowing for a gauge-invariant way of treating the light-matter interaction. In particular, expressing the position operator in the basis of WFs (dipole matrix elements) provides a straightforward path to computing the velocity matrix elements within the minimal-coupling scheme~\cite{yates_spectral_2007}. Furthermore, keeping track of the dipole matrix elements in the localized WFs basis also allows for performing the Power-Zienau-Woolley (PZW) transformation to the dipole gauge, which corresponds to a multi-orbital extension of the Peierls substitution~\cite{golez_multiband_2019,li_electromagnetic_2020-1,mahon_microscopic_2019,murakami_collective_2020}. The equivalence of velocity and dipole gauges for first-principle TB models has been investigated in our recent work~\cite{schuler_gauge_2021}.

In this work, the extend the theory to pump-probe trARPES, where the explicit time dependence of the strong THz pump pulse has to be taken into account. 
This paper is organized as follows. We discuss the light-matter coupling of the pump pulse to the sample in Sec.~\ref{subsec:lm_sample}, introducing the velocity and the dipole gauge from the perspective of (first-principle) WFs. For a self-contained presentation we briefly introduce the td-NEGF formalism in Sec.~\ref{sec:green}, which sets the stage for revisiting the open-system approach to photoemission in both gauges. Finally, we present simulated pump-probe spectra for monolayer graphene in Sec.~\ref{sec:examples}. Throughout this paper we use atomic units (a.u.) unless stated otherwise.

\section{Light-matter interactions in the sample\label{subsec:lm_sample}}

Before presenting the formalism of pump-probe trARPES, let us discuss how to incorporate light-matter interaction within the sample. A detailed discussion of the different gauges is presented in our work~\cite{schuler_gauge_2021}. For a self-contained presentation, we recapitulate the major points.

For a microscopic description, let us start from the minimal coupling principle. For clarity we consider a basis of (effectively) noninteracting bands, as obtained from density functional theory (DFT). In real space, the time-dependent Hamiltonian reads
\begin{align}
  \label{eq:ham_mincoup}
  \hat{h}(t) = \frac12 (\hat{\vec{p}}-q \vec{A}(t))^2 + v(\vec{r}) \ ,
\end{align}
where $q=-e$ is the charge of an electron, and $v(\vec{r})$ denotes the periodic potential (Kohn-Sham potential in the context of DFT). In this paper we use the dipole approximation, thus assuming the vector potential $\vec{A}(t)$ to not exhibit any spatial dependence. 

\subsection{Velocity gauge}

Let us now express all operators in a reduced basis spanned by the Bloch states $|\psi_{\vec{k}\alpha}\langle$, where the band index $\alpha$ runs over the relevant states. In the band basis, the time-dependent Hamiltonian $h_{\alpha \alpha^\prime}(\vec{k},t) = \langle \psi_{\vec{k}\alpha} | \hat{h}(t) | \psi_{\vec{k}\alpha^\prime} \rangle$ reads 
\begin{align}
    \label{eq:sp_ham_velo}
    h_{\alpha\alpha^\prime}(\vec{k},t) = \en_\alpha(\vec{k})\delta_{\alpha \alpha^\prime}
    -q \vec{A}(t)\cdot\vec{v}_{\alpha\alpha^\prime}(\vec{k}) + \frac{q^2}{2} \vec{A}(t)^2\delta_{\alpha \alpha^\prime} \ .
\end{align}
Here, the last term denotes the diamagnetic coupling, which reduces to a pure phase factor in the dipole approximation.
In Eq.~\eqref{eq:sp_ham_velo} we have introduced the velocity matrix elements 
\begin{align}
    \label{eq:velo_elemk}
    \vec{v}_{\alpha\alpha^\prime}(\vec{k}) &= \langle \psi_{\vec{k}\alpha} | \hat{\vec{p}} | \psi_{\vec{k}\alpha^\prime} \rangle = -i \langle \psi_{\vec{k}\alpha} | [\hat{\vec{r}},\hat{h}] | \psi_{\vec{k}\alpha^\prime} \rangle  \ .
\end{align}
Calculating the velocity matrix elements via Eq.~\eqref{eq:velo_elemk} is possible and often employed in first-principle calculations. However, for a clear physical interpretation and for computational advantages it is convenient to express Eq.~\eqref{eq:velo_elemk} as
\begin{align}
    \label{eq:velo_elemk_2}
    \vec{v}_{\alpha\alpha^\prime}(\vec{k}) = \nabla_{\vec{k}} \en_\alpha(\vec{k})\delta_{\alpha \alpha^\prime} - i \left(\en_{\alpha^\prime}(\vec{k}) - \en_{\alpha}(\vec{k})\right) \vec{A}_{\alpha\alpha^\prime}(\vec{k}) \ ,
\end{align}
where, $\vec{A}_{\alpha\alpha^\prime}(\vec{k}) = i \langle u_{\vec{k}\alpha} | \nabla_{\vec{k}}u_{\vec{k}\alpha^\prime} \rangle$ denotes the Berry connection. Calculating the velocity matrix elements by Eq.~\eqref{eq:velo_elemk_2} is most efficiently done by representing the Bloch wave-functions by localized WFs,
\begin{align}
    \label{eq:bloch_basis}
     | \psi_{\vec{k}\alpha} \rangle = \frac{1}{\sqrt{N}}\sum_{\vec{R}}e^{i \vec{k}\cdot\vec{R}}\sum_{m}C_{m\alpha}(\vec{k})
     |m\vec{R}\rangle \ .
\end{align}
Following ref.~\cite{yates_spectral_2007}, the Berry connection can be obtained from 
\begin{align}
    \label{eq:berryconect_2}
    \vec{A}_{\alpha\alpha^\prime}(\vec{k}) = \sum_{m m^\prime} C^*_{m\alpha}(\vec{k}) \left[
    \vec{D}_{m m^\prime}(\vec{k}) + i \nabla_{\vec{k}}\right]C_{m^\prime\alpha^\prime}(\vec{k}) \ .
\end{align}
Here we have defined the Fourier-transformed dipole operator
\begin{align}
    \label{eq:dipole_op}
    \vec{D}_{m m^\prime}(\vec{k}) = \sum_{\vec{R}} e^{i\vec{k}\cdot \vec{R}} \vec{D}_{m 0 m^\prime \vec{R}} \ ,
\end{align}
where $\vec{D}_{m \vec{R} m^\prime \vec{R}^\prime} = \langle m \vec{R} | \vec{r} - \vec{R} | m^\prime \vec{R}^\prime\rangle$ define the cell-centered dipole matrix elements. The $\vec{k}$-derivative in Eq.~\eqref{eq:berryconect_2} can be evaluated from an equivalent sum-over-states expression~\cite{yates_spectral_2007}. 

\subsection{Dipole gauge\label{subsec:dipgauge}}

Interpreting the Hamiltonian~\eqref{eq:ham_mincoup} as a finite system for a moment, the PZW transformation is defined by $\hat{U}(t) = \exp[-i q \vec{A}(t)\cdot \vec{r}]$. Applying this transformation to Eq.~\eqref{eq:ham_mincoup} yields the dipole gauge, where the light-matter coupling now has the form $\hat{h}_\mathrm{LM}(t) = -q \vec{E}(t)\cdot \vec{r}$, where $\vec{E}(t) = - d \vec{A}(t)/ dt$ is the electric field. The dipole operator $\vec{r}$ is ill-defined for periodic systems, which poses some technical difficulties. However, switching to the basis of localized WFs allows for a straightforward extension to periodic crystals. Hence, we introduce the Wannier Hamiltonian 
\begin{align}
  \label{eq:wann_ham}
  h_{m \vec{R} m^\prime \vec{R}^\prime} = \langle m \vec{R} | \frac{\hat{\vec{p}}^2}{2} + \hat{v} | m^\prime \vec{R}^\prime\rangle \ .
\end{align}
Translational invariance implies $h_{m \vec{R} m^\prime \vec{R}^\prime} = h_{m 0 m^\prime \vec{R}^\prime - \vec{R}}$, which connects to momentum space by 
\begin{align}
  h_{m m^\prime}(\vec{k}) = \sum_{\vec{R}} e^{i \vec{k} \cdot \vec{R}} h_{m 0 m^\prime \vec{R}} \ ,
\end{align}

The key idea is to define the PZW transformation in the space of WFs relative to the lattice sites as $U_{m \vec{R} m^\prime \vec{R}^\prime}(t) = \langle m \vec{R} | e^{-i  \vec{A}(t)\cdot(\vec{r} - \vec{R})} | m^\prime \vec{R}^\prime \rangle$~\cite{golez_multiband_2019,mahon_microscopic_2019-1}. As detailed in ref.~\cite{schuler_gauge_2021}, applying this time-dependent transformation and switching back to momentum space yields
\begin{align}
  \label{eq:ham_dip}
  \widetilde{h}_{m m^\prime}(\vec{k}, t) = h_{m m^\prime}(\vec{k} - q \vec{A}(t)) - q \vec{E}(t)\cdot \vec{D}_{m m^\prime}(\vec{k} - q \vec{A}(t)) \ .
\end{align}
Eq.~\eqref{eq:ham_dip} is a multi-orbital generalization of the Peierls substitution that captures intraband dynamics as well as local dipole transitions. 

Calculations can be performed in either the velocity (Eq.~\eqref{eq:sp_ham_velo}) or in the dipole gauge (Eq.~\eqref{eq:ham_dip}), and the gauge invariance of observables is guaranteed if the space of WFs forms a complete basis. This is strictly speaking not possible, as highly-excited Bloch states can not be presented by localized orbitals. Retaining approximate gauge invariance thus becomes a practical problem. As demonstrated for typical 2D systems in ref.~\cite{schuler_gauge_2021}, if the pump-induced dynamics is restricted to a set of bands spanned by well-localized WFs, observables obtained for either gauge are in excellent agreement.

\section{Many-body treatment from nonequilibrium Green's functions\label{sec:green}}

\begin{figure}[t]
\centering
\includegraphics[width=0.7\columnwidth]{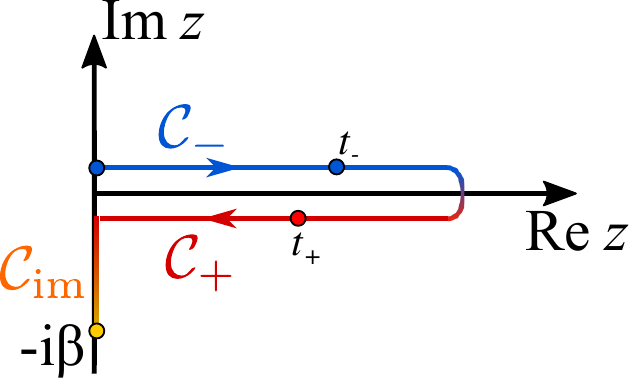}
\caption{The Kadanoff-Baym contour $\mathcal{C}$, running from $z=i0^+$ over the forward branch $\mathcal{C}_-$ to the backward branch $\mathcal{C}_+$, and through the imaginary branch $\mathcal{C}_\mathrm{im}$ to $z=-i \beta$.  \label{fig:contour}}
\end{figure}

Let us now present the formalism for treating many-body effects. We consider the Hamiltonian
\begin{align}
	\label{eq:hammb_velo}
	\hat{H}_\mathrm{VG}(t) = \sum_{\alpha\alpha^\prime} h_{\alpha \alpha^\prime}(\vec{k},t) \hcd_{\vec{k}\alpha} \hc_{\vec{k}\alpha^\prime} + \hat{H}_\mathrm{int}
\end{align}
in the velocity gauge, and
\begin{align}{}
	\label{eq:hammb_dip}
	\hat{H}_\mathrm{DG}(t) = \sum_{m m^\prime} \widetilde{h}_{m m^\prime}(\vec{k},t) \hcd_{\vec{k}m} \hc_{\vec{k}m^\prime} 
	 + \hat{H}^\prime_\mathrm{int} \ .
\end{align}
in the dipole gauge, where we have inserted the single-particle Hamiltonian~\eqref{eq:sp_ham_velo} and \eqref{eq:ham_dip}, respectively. We denote the electron creation (annihilation) operator by $\hat{c}^\dagger_{\vec{k} \ast}$ ($\hat{c}_{\vec{k} \ast}$).
All interaction effects are captured by $\hat{H}_\mathrm{int}$ ($\hat{H}^\prime_\mathrm{int}$). Since the PZW transformation is a purely spatial operator, any interaction derived from a spatial function is invariant under the transformation.
In particular, for a Coulomb-type interaction of the form $V(\vec{r}, \vec{r}^\prime)$, it is straightforward to show $\hat{H}_\mathrm{int} = \hat{H}^\prime_\mathrm{int}$.

\subsection{Equations of motion}

To capture the pump-induced dynamics including interaction effects, we introduce the velocity-gauge single-particle Green's functions (GFs)
\begin{align}
	\label{eq:gf_velo}
	G_{\alpha \alpha^\prime}(\vec{k}; z, z^\prime) = - i \langle T_{\mathcal{C}} \hc_{\vec{k}\alpha}(z) \hcd_{\vec{k}\alpha^\prime}(z^\prime) \rangle 
\end{align}
and the dipole-gauge GF
\begin{align}
	\label{eq:gf_dip}
	\widetilde{G}_{m m^\prime}(\vec{k}; z, z^\prime) = - i \langle T_{\mathcal{C}} \hc_{\vec{k}m}(z) \hcd_{\vec{k}m^\prime}(z^\prime) \rangle \ .
\end{align}
Here, $z, z^\prime$ denote arguments on the Kadanoff-Baym contour $\mathcal{C}   $(Fig.~\ref{fig:contour}), which conveniently includes the finite-temperature state and the real-time dynamics in the same formalism~\cite{stefanucci_nonequilibrium_2013}. The contour time evolution of the annihilation and creation operators is determined by the Hamiltonian~\eqref{eq:hammb_velo} for the velocity-gauge GF~\eqref{eq:gf_velo}, while the operators of the dipole-gauge GF~\eqref{eq:gf_dip} evolve with respect to Eq.~\eqref{eq:hammb_dip}. The symbol $T_\mathcal{C}$ denotes the contour-ordering operator, defined by the ordering indicated by the arrows in Fig.~\ref{fig:contour}.

The single-particle GF obeys an equation of motion on $\mathcal{C}$, which can be closed by introducing the self-energy, which captures all interaction effects. One obtains the Kadanoff-Baym equation (KBE) on the contour~\cite{stefanucci_nonequilibrium_2013}:
\begin{align}
	\label{eq:contour_kb}
	\left(i \partial_z - \vec{h}(\vec{k},z)\right)\vec{G}(\vec{k}; z, z^\prime) &= \delta_\mathcal{C}(z,z^\prime) \nonumber \\ 
	&+ \int_\mathcal{C} d\bar{z}\, \gvec{\Sigma}(\vec{k}; z, \bar{z}) \vec{G}(\vec{k}; \bar{z}, z^\prime) \ ,
\end{align}
where boldface symbols indicate a compact matrix notation in the band or orbital space. Eq.~\eqref{eq:contour_kb} describes the time evolution of the velocity-gauge GF~\eqref{eq:gf_velo}; the corresponding Kadanoff-Baym equation in the dipole gauge is obtained by replacing $\vec{h}\rightarrow \widetilde{\vec{h}}$, $\vec{G}\rightarrow \widetilde{\vec{G}}$, and $\gvec{\Sigma}\rightarrow \widetilde{\gvec{\Sigma}}$.

For both practical reasons and for a physical picture, the contour equation of motion~\eqref{eq:contour_kb} is usually solved by projecting onto observable times~\cite{schuler_nessi_2020}. For instance, the combination $z=t \in \mathcal{C}_0$, $z^\prime = t^\prime \in \mathcal{C}_+$  (see Fig.~\ref{fig:contour}) yields the lesser GF $G^<_{\alpha \alpha^\prime}(\vec{k}; t,t^\prime) = i \langle \hcd_{\vec{k}\alpha^\prime}(t) \hc_{\vec{k} \alpha}(t^\prime) \rangle$. The lesser GF contains information on the density matrix, $\rho_{\alpha \alpha^\prime}(\vec{k},t) = -i G^<_{\alpha^\prime \alpha}(\vec{k}; t, t)$ as well as on the time-dependent photocurrent, as discussed in Sec.~\ref{sec:trpes}. 

\subsection{Gauge transformation of Green's functions}

By solving the KBE~\eqref{eq:contour_kb} in either velocity or dipole gauge, we obtain the corresponding GF. How do the GFs related to each other? In absence of the pump field $\vec{A}(t) = 0$, $\vec{E}(t) = 0$, the single-particle Hamiltonians~\eqref{eq:sp_ham_velo} and \eqref{eq:ham_dip} are connected by a simple basis transformation,
\begin{align}
	\label{eq:ham_trans}
	\widetilde{h}_{m m^\prime} (\vec{k}) = h_{m m^\prime} = \sum_{\alpha \alpha^\prime} C_{m \alpha}(\vec{k}) h_{\alpha \alpha^\prime}(\vec{k}) C^*_{m^\prime \alpha^\prime}(\vec{k}) \ ,
\end{align}
where $C_{m \alpha}(\vec{k})$ are the coefficients relating band and WFs space (cf.~\eqref{eq:bloch_basis}), and $h_{\alpha \alpha^\prime} = \en_\alpha(\vec{k}) \delta_{\alpha \alpha^\prime}$. The basis transformation~\eqref{eq:ham_trans} can directly be applied to the annihilation and creation operators, i.\,e. $\hc_{\vec{k}m} = \sum_\alpha C_{m \alpha}(\vec{k}) \hc_{\vec{k}\alpha}$. Hence, the GF transforms as
\begin{align}
	\label{eq:gf_trans}
	\widetilde{G}_{m m^\prime}(\vec{k}; z, z^\prime) = \sum_{m m^\prime} C_{m \alpha}(\vec{k}) G_{\alpha \alpha^\prime}(\vec{k}; z, z^\prime) C^*_{m^\prime \alpha^\prime}(\vec{k}) \ .
\end{align}
This also implies that the density matrix transforms accordingly. In particular, the for band occupation $n_{\vec{k}\alpha}(t) = -i G^<_{\alpha \alpha}(\vec{k}; t, t^\prime)$ we find
\begin{align}
	\label{eq:band_occ}
	n_{\vec{k}\alpha}(t) = \widetilde{n}_{\vec{k}\alpha}(t) \equiv -i\sum_{m m^\prime}C^*_{m \alpha}(\vec{k}) \widetilde{G}^<_{m m^\prime}(\vec{k}; t,t) C_{m^\prime \alpha}(\vec{k})  \ .
\end{align}

In presence of the pump field $\vec{A}(t)$, this direct correspondence is lost, as Eq.~\eqref{eq:gf_trans} is broken. While observable quantities are gauge invariant, momentum-dependent occupations are not as Eq.~\eqref{eq:band_occ} is violated. Similarly, the lesser GF -- which plays the major role in theory of trARPES -- in the velocity gauge ($G^<_{\alpha \alpha^\prime}(\vec{k}; t,t ^\prime)$) and in the dipole gauge ($\widetilde{G}^<_{m m^\prime}(\vec{k}; t,t ^\prime)$) are no longer related by a unitary transformation in momentum space. Nevertheless, a generalized gauge transformation can be defined in space of WFs, as detailed in Appendix~\ref{app:gf_trans_gen}. To ensure the gauge invariance of the photocurrent -- which is an observable quantity -- we revisit the theory of trARPES from the td-NEGF perspective in both the velocity and the dipole gauge in the next section. 

\section{Time-resolved photoemission from embedding theory\label{sec:trpes}}

\begin{figure}[t]
\centering
\includegraphics[width=\columnwidth]{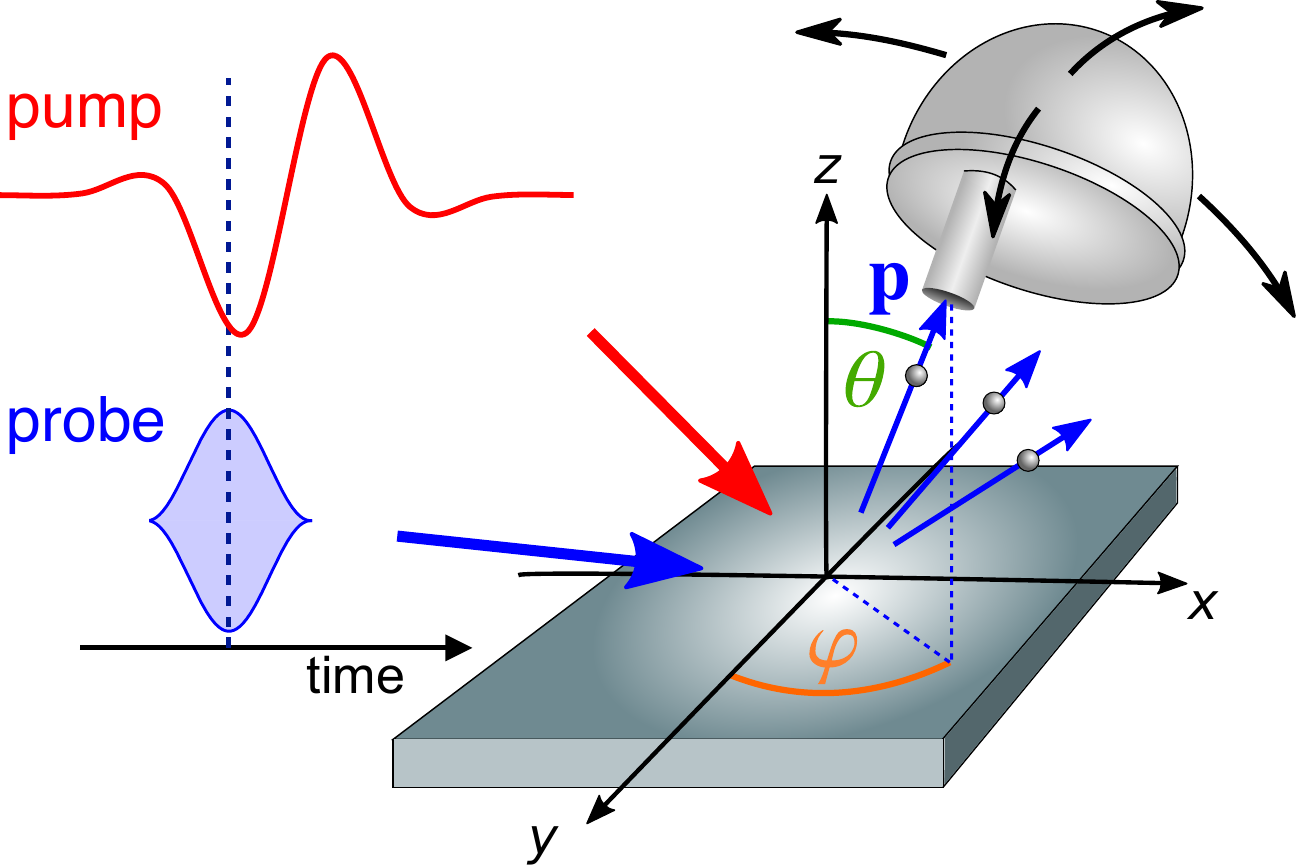}
\caption{Setup of pump-probe trARPES: the delay of the pump and the probe pulse is varied on a subcycle time scale. The momentum of the photoelectrons is denoted by $\vec{p}$, whose spherical coordinates are described by the angles $(\theta,\varphi)$. \label{fig:setup}}
\end{figure}

Let us introduce the geometry and experimental setup we are modelling in this paper, sketched in Fig.~\ref{fig:setup}.
While the THz pump pulse induces excitations and transiently dresses the electronic structure, the probe pulse (we assume a typical XUV pulse) photoemits the electrons with momentum $\vec{p}$, which are captured by a detector. We assume strength of the probe pulse to be in the perturbative regime. In absence of the pump pulse, fixing the acceptance energy $\en_f$ and measuring the emission angle $(\theta, \varphi)$ yields the three-dimensional momentum $\vec{p} = \sqrt{2 \en_f}(\cos\varphi\sin\theta, \sin\varphi\sin\theta, \cos\theta)$. We assume the same map $(\en_f,\theta,\varphi) \rightarrow \vec{p}$ also in presence of the pump pulse it is fixed by the detector. Note that the pump pulse modifies the energy of the photoelectron (streaking effects).

The sample is assumed to be periodic two-dimensional (2D) material. This treatment naturally also includes bulk materials, as the probe pulse penetrates only through a small number of{} atomic layers. The out-of-plane crystal momentum $k_z$ (which is not conserved in the photoemission process) can be included in the set of band indices. In contrast, the in-plane crystal momentum $\vec{k}$ is conserved up to a reciprocal lattice vector $\vec{G}$. This momentum conservation is naturally captured by the photoemission matrix elements between the Bloch states and the photoelectron states $|\chi_\vec{p}\rangle$:
\begin{align}
	\label{eq:pes_mel}
	M_\alpha(\vec{k},\vec{p}) = \langle \chi_{\vec{p}} | \vec{e} \cdot \hat{\vec{p}} | \psi_{\vec{k}\alpha} \rangle = \delta_{\vec{p}_\parallel, \vec{k} + \vec{G}}  \langle \chi_{(\vec{k},p_\perp)} | \vec{e} \cdot \hat{\vec{p}} | \psi_{\vec{k}\alpha} \rangle  \ .
\end{align}
Here, $\vec{e}$ denotes the polarization of the probe pulse. Fixing $\vec{k}$ by the detector angles, the out-of-plane momentum of the photoelectron $p_\perp$ is determined from the energy $\en_f$ by $2\en_f = (\vec{k}+\vec{G})^2 + p^2_\perp$. 

\subsection{Photocurrent in the velocity gauge \label{subsec:photo_velo}}

To derive the photocurrent (driven by the probe pulse), we extend the Hamiltonian~\eqref{eq:hammb_velo} by the subspace of photoelectrons:
\begin{align}
	\label{eq:ham_ext_velo}
	\hat{H}^\mathrm{ext}_\mathrm{VG}(t) &= \hat{H}_\mathrm{VG}(t) - \left\{q A(t)\sum_{\vec{k}\alpha} \sum_{\vec{p}} M_\alpha(\vec{k},\vec{p}) \hat{d}^\dagger_\vec{p} \hc_{\vec{k}\alpha} + \mathrm{h.c.} \right\} \nonumber \\ &+
	\sum_{\vec{p},\vec{p}^\prime} \langle \chi_\vec{p} | \hat{h}(t) | \chi_{\vec{p}^\prime} \rangle \hat{d}^\dagger_\vec{p} \hat{d}_{\vec{p}^\prime } \ .
\end{align}
Here, $\vec{A}(t) = \vec{e} A(t)$, while $\hat{d}^\dagger_\vec{p}$ ($\hat{d}_\vec{p}$) stands for the creation (annihilation) operator with respect to the photoelectron state $|\chi_{\vec{p}}\rangle$. In principle, the laser field entering the Hamiltonian~\eqref{eq:ham_mincoup} can give rise off-diagonal coupling among the photoelectron states.  Especially for molecules, this effect is known as Coulomb-laser coupling and plays an import role for accurately modeling pump-probe spectrograms~\cite{pazourek_time-resolved_2013}. Here we assume that all interactions are sufficiently screened and that the photoelectron energy is large, allowing us to approximate
\begin{align}
	\label{eq:chip_diag}
	 \langle \chi_\vec{p} | \hat{h}(t) | \chi_{\vec{p}^\prime} \rangle &\approx \left(\frac{\vec{p}^2}{2} - q \langle \chi_{\vec{p}} | \vec{A}(t) \cdot \hat{\vec{p}} | \chi_{\vec{p}} \rangle  + \frac{q^2 \vec{A}(t)^2}{2}\right)\delta_{\vec{p}\vec{p}^\prime} \nonumber 
	 \\ &\equiv \en_{\vec{p}}(t)\delta_{\vec{p}\vec{p}^\prime}
\end{align}

The vector potential entering the Hamiltonian~\eqref{eq:ham_ext_velo} contains both the pump and the probe pulse: $\vec{A}(t) = \vec{A}_\mathrm{p}(t) + \vec{A}_\mathrm{pr}(t)$. We assume $|\vec{A}_\mathrm{pr}(t)| \ll |\vec{A}_\mathrm{p}(t)|$, and treat all effects driven by the probe pulse as small perturbation. In this scenario, only the probe pulse can give rise to photoemission. Hence, we replace $A(t)\rightarrow A_\mathrm{pr}(t)$ in the second term in Eq.~\eqref{eq:ham_ext_velo}, while all other terms only include the pump field. Similar as in refs.~\cite{freericks_theoretical_2009,perfetto_pump-driven_2019,perfetto_time-resolved_2020}, we can compute the photocurrent as the flow of electrons into the photoelectron space. We describe this effect by an embedding self-energy, which is defined by
\begin{align}
	\label{eq:sigma_emb}
	\Sigma^\mathrm{R}_{\alpha \alpha^\prime}(\vec{k}; t,t^\prime) = A^*_\mathrm{pr}(t)A_\mathrm{pr}(t^\prime) M^*_\alpha(\vec{k},\vec{p}) M_{\alpha^\prime}(\vec{k},\vec{p}) g^\mathrm{R}_{\vec{p}}(t,t^\prime) \ ,
\end{align}
where 
\begin{align}
	\label{eq:gret_pe}
	g^\mathrm{R}_{\vec{p}}(t,t^\prime) = -i \theta(t-t^\prime)\exp\left(-i \int^t_{t^\prime} d\bar{t} \en_{\vec{p}}(\bar{t})\right)
\end{align}
is the retarded GF of the photoelectrons. Now employing the transient Meier-Wingreen formula~\cite{stefanucci_nonequilibrium_2013}, time-dependent photocurrent reads
\begin{align}
	\label{eq:dotnp}
	\dot{N}_{\vec{p}}(t) = \mathrm{Re}\sum_{\vec{k}}\sum_{\alpha \alpha^\prime} \int^t_0 d t^\prime \Sigma^\mathrm{R}_{\alpha \alpha^\prime}(\vec{k}; t,t^\prime) G^<_{\alpha^\prime \alpha}(\vec{k}; t^\prime, t) \ ,
\end{align}
where we have used $g^<_{\vec{p}}(t,t^\prime) = 0$ since there are no photoelectrons present in equilibrium.
In a pump-probe setup the photoelectrons are detected over a long time interval $T\rightarrow \infty$, and the trARPES signal is given by $I(\vec{p}) = (1/T) \int^T_0 dt \dot{N}_{\vec{p}}(t)$. Parameterizing the probe pulse as $A_\mathrm{pr}(t) = A_0 s(t) e^{i \omega_\mathrm{pr} t}$ with the pulse envelop $s(t)$, combining Eq.~\eqref{eq:sigma_emb}--\eqref{eq:dotnp} yields 
\begin{widetext}
\begin{align}
	\label{eq:trarpes_velo}
	I(\vec{p}) \propto \mathrm{Im}\sum_{\vec{k}} \sum_{\alpha \alpha^\prime} M^*_\alpha(\vec{k},\vec{p}) M_{\alpha^\prime}(\vec{k},\vec{p}) \int^\infty_0 \!dt\!\!\int^t_0 \! d t^\prime 
	s(t)s(t^\prime)\exp\left(-i \int^t_{t^\prime} d\bar{t} \left[\en_{\vec{p}}(\bar{t}) -\omega_\mathrm{pr} \right]\right)
	G^<_{\alpha^\prime \alpha}(\vec{k}; t^\prime, t) \ .
\end{align}
\end{widetext}
Eq.~\eqref{eq:trarpes_velo} is a generalization of the expression known from the literature~\cite{freericks_theoretical_2009,sentef_examining_2013}, where the laser-dressing of the continuum states -- known as laser-assisted photoemission (LAPE)~\cite{miaja-avila_laser-assisted_2006} -- is taken into account. Note that by construction $\dot{N}_\vec{p}(t) \ge 0$, implying $I(\vec{p}) \ge 0$, which is a fundamental requirement for a gauge-invariant description of trARPES~\cite{freericks_gauge_2015}.

\subsection{Photocurrent in the dipole gauge \label{subsec:photo_dip}}

Before repeating the analogous steps as in Sec.~\ref{subsec:photo_velo}, we need to transform the extended Hamiltonian~\eqref{eq:ham_ext_velo} to the dipole gauge. This is accomplished by introducing the unitary transformation discussed in Sec.~\ref{subsec:dipgauge} on a many-body level as $\hat{U}(t) = e^{\hat{S}(t)}$ with
\begin{align}
	\label{eq:sgen}
	\hat{S}(t) = -i q\vec{A}(t)\cdot \sum_{\vec{R},\vec{R}^\prime} \sum_{m m^\prime} \vec{D}_{m \vec{R} m^\prime \vec{R}^\prime} \hcd_{m \vec{R}} \hc_{m^\prime \vec{R}^\prime} \ .
\end{align}
Applying the transformation defined by the generator~\eqref{eq:sgen} to the Hamiltonian~\eqref{eq:ham_ext_velo} defines the extended dipole-gauge Hamiltonian. The band space without the photoelectrons transforms (as captured by the Hamiltonian~\eqref{eq:hammb_velo}) according to the single-particle picture as discussed in Sec.~\ref{subsec:dipgauge}, i.\, e. $\hat{H}_\mathrm{DG}(t) = \hat{U}(t) \hat{H}_\mathrm{VG}(t) \hat{U}^\dagger(t) + i \partial_t \hat{S}(t)$, where $\hat{H}_\mathrm{DG}(t)$ is identical to Eq.~\eqref{eq:hammb_dip}.

Applying the transformation~\eqref{eq:sgen} to the photoelectron subspace is not possible. Delocalized continuum states can not be represented by localized WFs, which renders the dipole matrix elements with respect to $|\chi_{\vec{p}} \rangle$ ill-defined. We thus assume that the orbital space defining the generator~\eqref{eq:sgen} does not include photoelectrons, and treat $\hat{U}(t)$ as identity transformation when acting on the photoelectron operators $\hat{d}_{\vec{p}}$. 
Using the approximation~\eqref{eq:chip_diag}, we obtain the dipole-gauge extended Hamiltonian 
\begin{align}
	\label{eq:ham_ext_dip}
	\hat{H}^\mathrm{ext}_\mathrm{DG}(t) = \hat{H}_\mathrm{DG}(t) + \sum_{\vec{p}} \en_{\vec{p}}(t) \hat{d}^\dagger_\vec{p} \hat{d}_{\vec{p}} + \hat{H}^\mathrm{pes}_\mathrm{DG}(t) \ ,
\end{align}
where
\begin{align}
	\label{eq:ham_pes_dip}
	\hat{H}^\mathrm{pes}_\mathrm{DG}(t) &= - q A_\mathrm{pr}(t)\sum_{\vec{k}\alpha} \sum_{\vec{p}} M_\alpha(\vec{k},\vec{p}) \hat{d}^\dagger_\vec{p} \hat{U}(t)\hc_{\vec{k}\alpha}
	 \hat{U}^\dagger(t) \nonumber \\ &\quad + \mathrm{h.c.}
\end{align}
is the photoemission term. The latter can be evaluated by to the WFs basis, as detailed in Appendix~\ref{app:trarpes_gauge}. In essence, expressing the unitary transformation $\hat{U}(t)$ by its action on the Bloch basis, the time dependence enters now time-dependent photoemission matrix elements
\begin{align}
	\label{eq:tdmel}
	\widetilde{M}_m(\vec{k},\vec{p},t) = \langle \chi_{\vec{p}} | \vec{e}\cdot \hat{\vec{p}} e^{i q\vec{A}(t)\cdot\vec{r}} | \phi_{\vec{k}-q \vec{A}(t)m} \rangle \ ,
\end{align}
where $|\phi_{\vec{k}m}\rangle = 1/\sqrt{N}\sum_{\vec{R}}e^{i \vec{k}\cdot \vec{R}}|m \vec{R}\rangle$ is the basis spanned by the WFs.
The coupling of band electrons to the photoelectrons thus attains the form
\begin{align}
	\label{eq:ham_pes_dip_fin}
	\hat{H}^\mathrm{pes}_\mathrm{DG}(t) &= - q A_\mathrm{pr}(t)\sum_{\vec{k}m} \sum_{\vec{p}} \widetilde{M}_m(\vec{k},\vec{p},t) \hat{d}^\dagger_\vec{p} \hc_{\vec{k}m}+ \mathrm{h.c.} \ .
\end{align}
Note that that time-dependent matrix elements~\eqref{eq:tdmel} ensure in-plane momentum conservation $\vec{k}+\vec{G} =\vec{p}_\parallel$, as detailed in Appendix~\ref{app:mom_dip}.

Starting from Eq.~\eqref{eq:ham_ext_dip} and \eqref{eq:ham_pes_dip_fin}, we can now follow the analogous steps as in Sec.~\ref{subsec:photo_velo} to derive the photocurrent from the embedding method. We obtain
\begin{widetext}
\begin{align}
	\label{eq:trarpes_dip}
	\widetilde{I}(\vec{p}) \propto \mathrm{Im}\sum_{\vec{k}} \sum_{m m^\prime} \int^\infty_0 \!dt\!\!\int^t_0 \! d t^\prime 
	s(t)s(t^\prime) \widetilde{M}^*_m(\vec{k},\vec{p},t) \widetilde{M}_{m^\prime}(\vec{k},\vec{p},t^\prime) \exp\left(-i \int^t_{t^\prime} d\bar{t} \left[\en_{\vec{p}}(\bar{t}) -\omega_\mathrm{pr} \right]\right)
	\widetilde{G}^<_{m^\prime m}(\vec{k}; t^\prime, t) \ .
\end{align}
\end{widetext}
The embedding formalism directly yields the measurable current, hence the trARPES intensity ~\eqref{eq:trarpes_dip} is non-negative. The explicit time dependence of the matrix elements~\eqref{eq:tdmel} can be intuitively understood by comparing the velocity (Eq.~\eqref{eq:sp_ham_velo}) and the dipole gauge Hamiltonian (Eq.~\eqref{eq:ham_dip}): the momentum space is shifted by $q\vec{A}(t)$ in the dipole gauge. This effect is compensated by the time-dependence of the matrix elements. Because Eq.~\eqref{eq:ham_ext_dip} and Eq.~\eqref{eq:ham_ext_velo} are related by a unitary transformation, gauge invariance of the trARPES signal $I(\vec{p}) = \widetilde{I}(\vec{p})$ is fulfilled by construction. In Appendix~\ref{app:trarpes_gauge} we show this equivalence explicitly.

\section{Application: subcycle photoemission from graphene \label{sec:examples}}

\begin{figure}[t]
\centering
\includegraphics[width=\columnwidth]{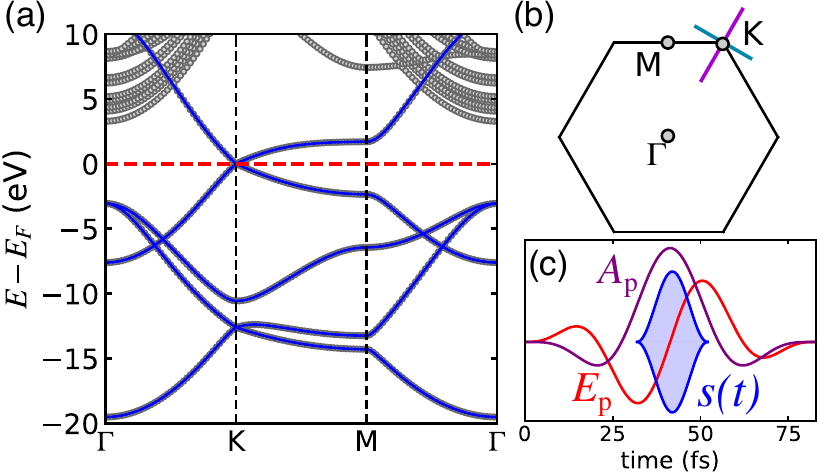}
\caption{(a) Calculated bands structure from DFT (circles) and the first-principle TB model (solid lines). (b) Brillouin zone of graphene where the high-symmetry points are marked. The lines close to K indicate the path in momentum space that we will focus on for presenting spectra. (c) Illustration of the electric field (vector potential) of the pump pulse $E_\mathrm{p}(t)$ ($A_\mathrm{p}(t)$) and the envelop $s(t)$ of the probe pulse. \label{fig:bands}}
\end{figure}

We apply the theory to time-resolved photoemission from a monolayer graphene. To construct a TB model from first principles, we performed DFT calculations with the {\sc Quantum espresso} code~\cite{giannozzi_quantum_2009}. The exchange and correlation effects are treated on the level of the local-density approximation (LDA). Norm-conserving pseudopotentials from the \textsc{PseudoDojo} project~\cite{van_setten_pseudodojo_2018} were used. The self-consistent DFT calculation was performed with a $16\times 16$ Monkhorst-Pack sampling of the Brillouin zone (BZ). We used the {\sc Wannier90} code~\cite{mostofi_updated_2014} to obtain projective WFs. The $sp_2$ hybridized and $p_z$ orbitals were taken as initial guess. This procedure yields the matrix elements of the Hamiltonian~\eqref{eq:wann_ham} and the dipole matrix elements $\vec{D}_{m \vec{R} m^\prime \vec{R}^\prime}$. As custom code is then used to compute the velocity matrix elements~\eqref{eq:velo_elemk_2}, defining the velocity-gauge Hamiltonian~\eqref{eq:sp_ham_velo}, while the dipole-gauge Hamiltonian~\eqref{eq:ham_dip} is directly available.
We solve the KBE~\eqref{eq:contour_kb} by using the NESSi code~\cite{schuler_nessi_2020}. Here we focus on the light-matter interaction specifically and neglect any correlation or scattering effects. After obtaining the lesser GF $G^<_{\alpha \alpha^\prime}(\vec{k}; t, t^\prime)$ and $\widetilde{G}^<_{m m^\prime}(\vec{k}; t, t^\prime)$, we compute the trARPES spectra via Eq.~\eqref{eq:trarpes_velo} and \eqref{eq:trarpes_dip}, respectively. 

\subsection{Photoemission matrix elements}

For computing the photoemission matrix elements we focus on the $\pi$ bands, thus including only the $p_z$ orbitals. The final states $|\chi_{\vec{p}}\rangle$ are treated within the plane-wave (PW) approximation. This approach simplifies the calculations significantly while capturing basic features like the dark corridor~\cite{gierz_illuminating_2011}, which is a region of low photoemission intensity due to destructive interference from the sublattice sites. Using the Wannier presentation~\eqref{eq:bloch_basis} the velocity-gauge matrix elements~\eqref{eq:pes_mel}
reduce to
\begin{align}
  \label{eq:mel_pw}
  M_{\alpha}(\vec{k},\vec{p}) = \delta_{\vec{p}_\parallel, \vec{k} + \vec{G}}\, \vec{e}\cdot \vec{p} \int d\vec{r}\, \sum_{m} C_{m\alpha}(\vec{k}) e^{-i\vec{p}\cdot\vec{r}}\phi_{p_z}(\vec{r}-\vec{t}_m) \ .
\end{align}
Here, the WF $\phi_{p_z}(\vec{r})$ is modeled by an hydrogen-like atomic orbitals with the corresponding angular momentum, centered at either of the two carbon sites (position $\vec{t_m}$) in the unit cell (similar as in ref.~\cite{schuler_local_2020-1}).

The time-dependent matrix elements within the dipole gauge~\eqref{eq:tdmel} are evaluated similarly, yielding
\begin{align}
  \label{eq:tdmel_pw}
  \widetilde{M}_m(\vec{k},\vec{p},t) = \delta_{\vec{p}_\parallel, \vec{k} + \vec{G}}\, \vec{e}\cdot \vec{p} \int d\vec{r}\,  e^{-i(\vec{p}-q\vec{A}(t))\cdot\vec{r}}\phi_{p_z}(\vec{r}-\vec{t}_m) \ .
\end{align}
Note the time-dependent shift in the PW function in Eq.~\eqref{eq:tdmel_pw}, which compensates for the gauge differences of the GFs. 

\begin{figure*}[t]
\centering
\includegraphics[width=\textwidth]{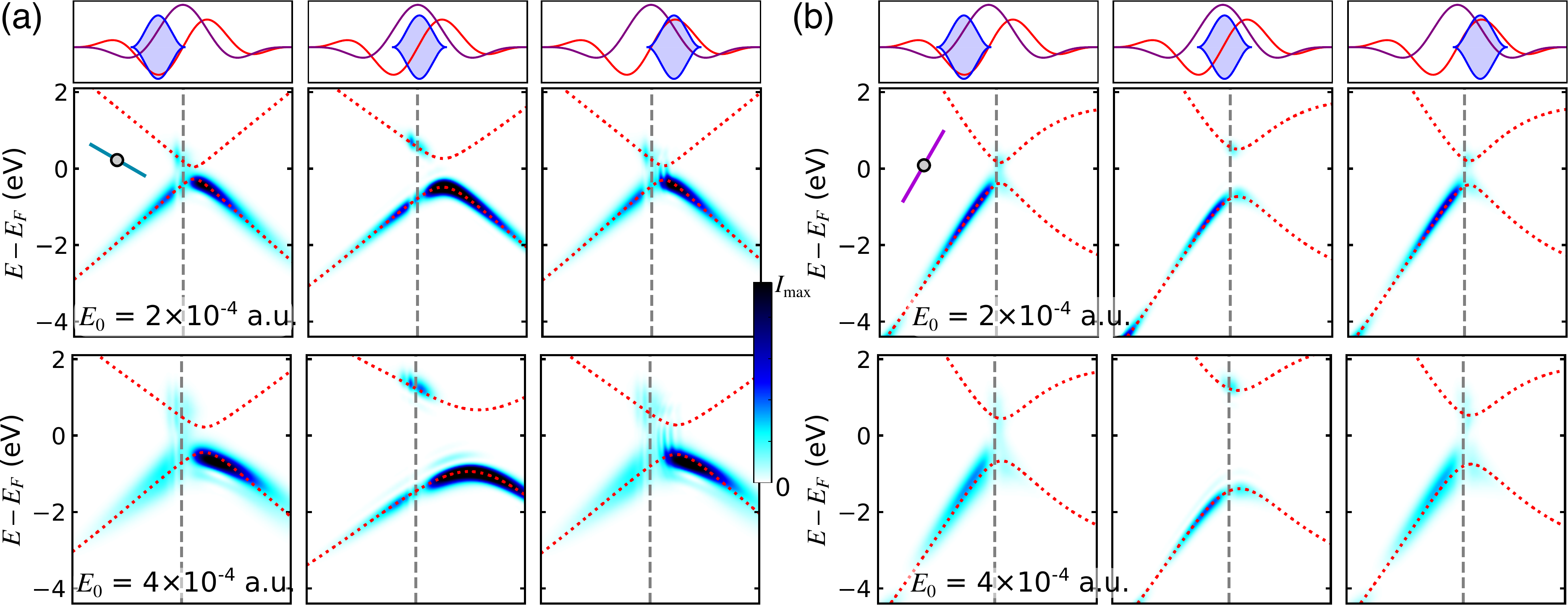}
\caption{(a) Pump-probe spectra along the path orthogonal to the $\Gamma$-K direction (see Fig.~\ref{fig:bands})(b) for different relative positions of the pump and the probe pulse (top row). The vertical dashed line indicates the position of K. The red dotted lines represent the averaged adiabatic band structure. Middle panels: $E_0 = 2\times 10^{-4}$  a.u., bottom panels: $E_0 = 4\times 10^{-4}$ a.u.. (b) Same as (a), but for the path along the $\Gamma$-K direction. All calculations were performed in the velocity gauge. \label{fig:spectra_velo}}
\end{figure*}

\subsection{Time-resolved spectra}

We simulated dynamics induced by a short THz pulse, parameterized by
\begin{align}
  \label{eq:pump}
  \vec{E}_\mathrm{p}(t) = \vec{e}_\mathrm{p} E_0 \sin^2\left(\frac{\omega_\mathrm{p}t}{2 n_c}\right) \sin(\omega_\mathrm{p} t) \ , \ 0 \le t \le \frac{2\pi}{\omega_\mathrm{p}}n_c \ ,
\end{align}
where $n_c$ is the number of optical cycles, which we fix to $n_c=2$. The pump polarization $\vec{e}_\mathrm{p}$ is chosen along the $x$-direction, while for the pump frequency we consider $\hbar \omega_\mathrm{p}=0.1$~eV. The photon energy of the probe pulse is fixed at the typical value $\hbar \omega_\mathrm{pr} = 22$~eV. We use the probe envelop $s(t) = \cos^2[\pi (t-\Delta t)/T_\mathrm{pr}]$ for $-T_\mathrm{pr}/2 \le t \le T_\mathrm{pr}/2$ with a duration of $T_\mathrm{pr}=20$~fs.

With these ingredients we computed the trARPES spectra along the paths in the BZ sketched in Fig.~\ref{fig:bands}(b). Here we opted for the velocity gauge. The spectra, calculated from Eq.~\eqref{eq:trarpes_velo}, are presented in Fig.~\ref{fig:spectra_velo}. For illustrating the effects we consider $E_0=2\times 10^{-4}$ a.u. (strong pulse) and $E_0=4\times 10^{-4}$ a.u. (very strong pulse). 

Inspecting the spectra in Fig.~\ref{fig:spectra_velo}(a) we notice a shift of the photodressed bands, especially when the probe pulse is centered at the maximum of the envelop of the pump~\eqref{eq:pump}, where the vector potential $\vec{A}_\mathrm{p}(t)$ reaches its maximum amplitude. An opening of the gap is also observed. In the regime of low pump frequency, this shift can be understood in the adiabatic picture. We calculated the band structure from Eq.~\eqref{eq:sp_ham_velo} assuming constant vector potential $\vec{A}_0(\Delta t) = \int d t \vec{A}_\mathrm{p}(t) s(t) / \int dt s(t)$, which is shown by the red dotted lines in Fig.~\ref{fig:spectra_velo}. This analysis demonstrates that the adiabatic photodressing of the bands is the predominant feature. Note that close to the Dirac point the time evolution can never be adiabatic due to the vanishing gap. Therefore, direct transitions occur, leading to the region of suppressed (enhanced) intensity in the lower (upper) band at K. Another interesting feature is the pronounced broadening observed when the probe pulse overlaps with the maximum or minimum of the pump electric field. During this time interval the vector potential varies the most, resulting in a broader distribution of instantaneous values. This effect scales with the pump field strength. Very similar effects are also observed for the path in the BZ along the $\Gamma$-K direction (Fig.~\ref{fig:spectra_velo}(b)). Since the path is mostly along the $k_y$-direction the adiabatic shift of the bands is less pronounced. 

\subsection{Velocity gauge vs. dipole gauge}

\begin{figure}[b]
\centering
\includegraphics[width=\columnwidth]{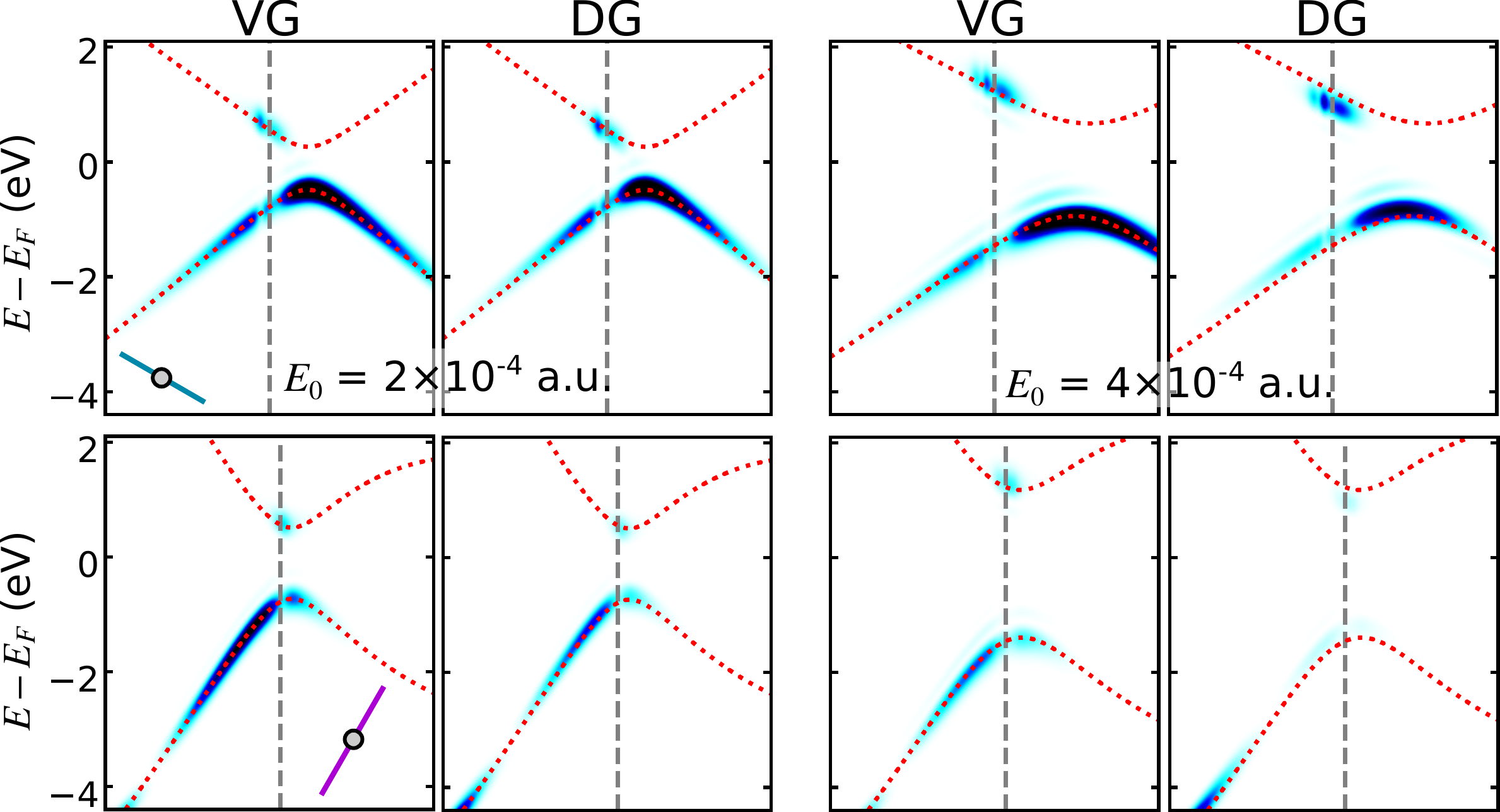}
\caption{Comparison of the velocity gauge (VG) and the dipole gauge (DG) for $E_0 = 2\times 10^{-4}$ a.u. (left four panels) and $E_0 = 4\times 10^{-4}$ a.u. (right four panels). The upper (lower) row corresponds to path in BZ shown in Fig.~\ref{fig:spectra_velo}(a) (Fig.~\ref{fig:spectra_velo}(b)).  \label{fig:compare}}
\end{figure}

The gauge invariance of the time-resolved spectra~\eqref{eq:trarpes_velo} and \eqref{eq:trarpes_dip} is guaranteed provided that the space of WFs is complete. Note that this is strictly speaking always an approximation, as including both localized and delocalized orbitals is required. In practice, the equivalence of the gauges for a given Wannier model has to be checked explicitly. Furthermore, the approximation of the WFs by hydrogen-like orbitals entering the matrix elements~\eqref{eq:mel_pw} and \eqref{eq:tdmel_pw} is in principle inconsistent with the Hamiltonian. 

Therefore, we have performed the analogous simulation of the trARPES signal within the dipole gauge, inserting the matrix elements~\eqref{eq:tdmel_pw}. 
The comparison of the spectra is shown in Fig.~\ref{fig:compare} for the probe pulse centered at the peak vector potential $\vec{A}_\mathrm{p}(t)$. For moderate pulse strength $E_0 = 2\times 10^{-4}$ a.u. the spectra within the velocity and the dipole gauge agree very well, especially perpendicular to the $\Gamma$-K direction. Both gauges correctly reproduce the photodressed band structure; there are small deviations in the strong-field regime. The more striking discrepancy is the reduction of intensity in the dipole gauge, in particular along the $\Gamma$-K direction. Note that matrix element effects are strongly pronounced in this direction, including the dark corridor. Inspecting other observables we find a quantitative deviation of the velocity and the dipole gauge, but the the inconsistency the trARPES intensity is mostly attributed to the approximate treatment of the photoemission matrix elements~\eqref{eq:tdmel_pw}. Inserting the WFs directly obtained from \textsc{Wannier90} is expected to result in better agreement.

\subsection{Influence of the probe pulse duration}

\begin{figure}[t]
\centering
\includegraphics[width=\columnwidth]{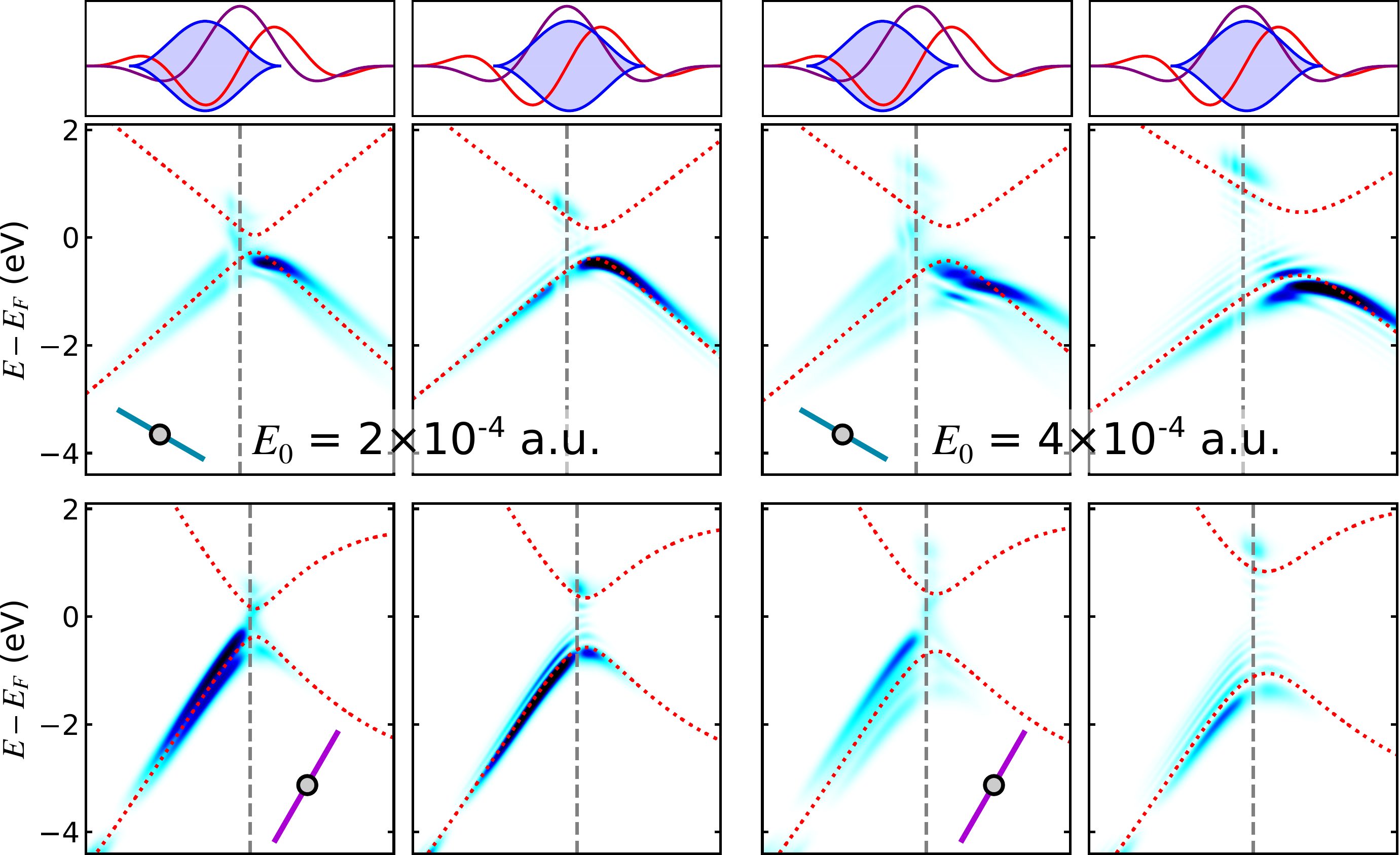}
\caption{Pump-probe spectra (calculated within the velocity gauge) for $E_0 = 2\times 10^{-4}$ a.u. (left four panels) and $E_0 = 4\times 10^{-4}$ a.u. (right four panels) for probe length $T_\mathrm{pr}=40$~fs. \label{fig:longpulse}}
\end{figure}

There are several features in the spectra in Fig.~\ref{fig:spectra_velo} attributed to the time-varying pump field $\vec{A}_\mathrm{p}(t)$. Apart from the pronounced broadening when the probe pulse overlaps with peaks of $\vec{E}_\mathrm{p}(t)$, a side band at appears roughly 200~meV above the main band when the probe pulse is centered at the maximum of $\vec{A}_\mathrm{p}(t)$. To explore both effects in more detail we increased the probe pulse duration to $T_\mathrm{pr} = 40$~fs and calculated the corresponding trARPES spectra (Fig.~\ref{fig:longpulse}) in the velocity gauge. 

The longer probe duration increases the frequency resolution, revealing a series of side bands, which are particularly visible for stronger $E_0$. These side bands can be traced back to the slowly varying pump field, which effectively chirps the oscillatory time dependence of $G^<_{\alpha\alpha^\prime}(\vec{k}; t, t^\prime)$. These effects can be explored by analyzing a simple model system. Let consider the GF $g^<(t,t^\prime) = -i \exp\{-i \lambda \int^t_{t^\prime} d\bar{t}\, \alpha(\bar{t}) \}$. The phase factor $\alpha(t) = \omega_\mathrm{p} A_\mathrm{p}(t) / E_0$ mimics the intraband coupling, the scale of which set by the parameter $\lambda$. The corresponding pump-probe spectrum is calculated by a simplified version of Eq.~\eqref{eq:trarpes_velo},
\begin{align}
    \label{eq:model_spectrum}
    I(\omega) = \mathrm{Im}\int^\infty_0\! dt \! \int^t_0\! d t^\prime \, s(t)s(t^\prime) e^{-i \omega (t-t^\prime)} g^<(t,t^\prime) \ .
\end{align}
In absence of the pump pulse ($\lambda = 0$), the spectrum~\eqref{eq:model_spectrum} is peak at $\omega=0$ with a broadening set by the probe duration $T_\mathrm{pr}$. Increasing $\lambda$ simulates the impact of the finite pump pulse.

\begin{figure}[t]
\centering
\includegraphics[width=\columnwidth]{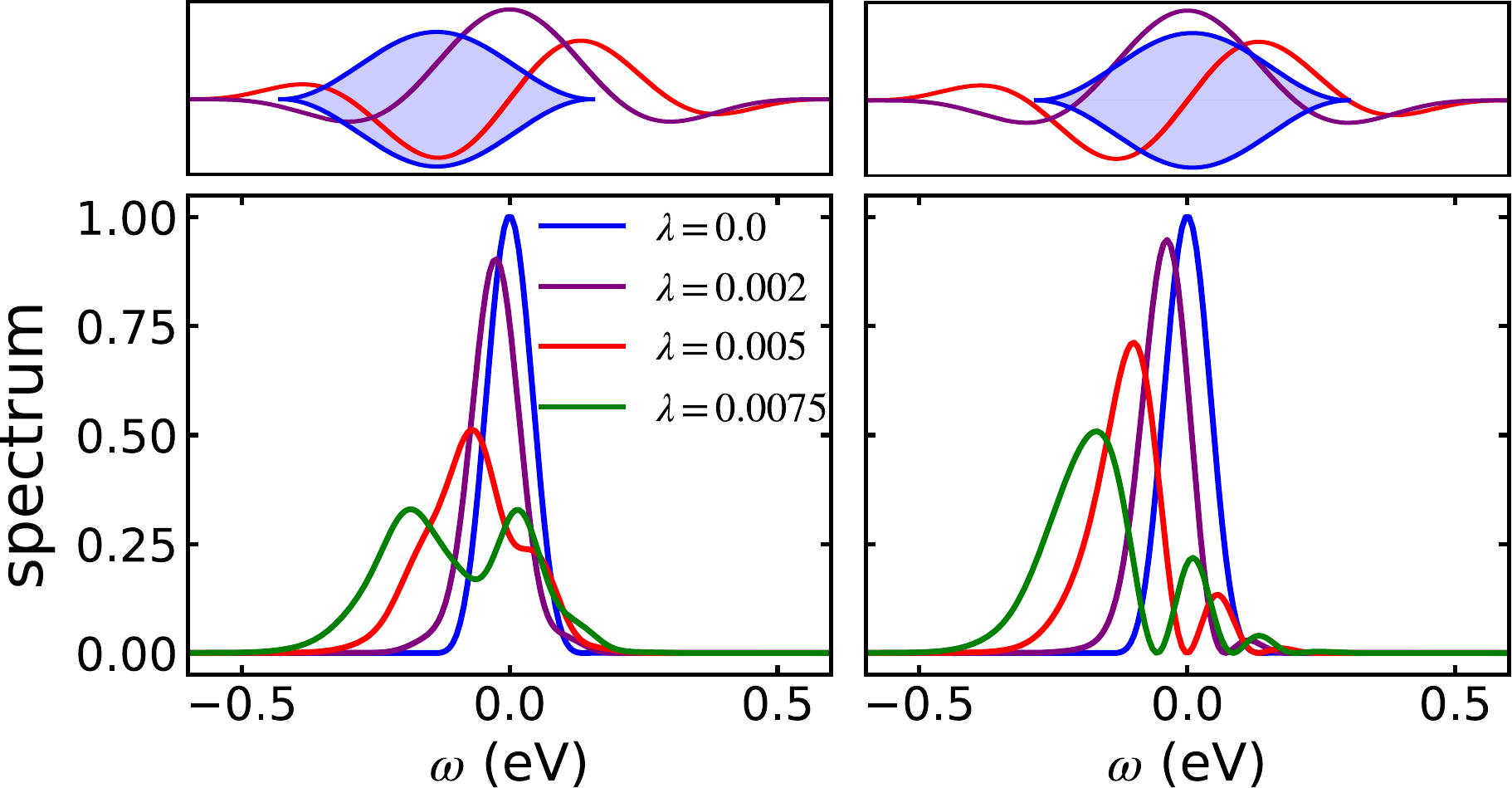}
\caption{Model pump-probe spectra~\eqref{eq:model_spectrum} for different values of the intraband coupling $\lambda$. The relative position of the probe pulse with $T_\mathrm{pr} = 40$~fs is sketched in the top panels. \label{fig:chirp}}
\end{figure}

The corresponding spectra, presented in Fig.~\ref{fig:chirp}, show a lot of qualitative resemblance with the simulated trARPES spectra of graphene in Fig.~\ref{fig:longpulse}. While for small $\lambda$ the spectrum $I(\omega)$ is slightly shifted and broadened, additional peaks appear for larger $\lambda$. The variation of $\alpha(t)$ over a larger interval in the left panel of Fig.~\ref{fig:chirp} gives rise to pronounced broadening, while multiple side peaks are apparent for $s(t)$ centered at the maximum vector potential. The analysis of this simplified model demonstrates that the side peak structure and broadening effects observed in Fig.~\ref{fig:spectra_velo}  and \ref{fig:longpulse} are due to the intraband acceleration during the probe pulse. 

\section{Conclusions}

We introduced the theory of time-resolved photoemission from the td-NEGF framework in the scenario where both the probe pulse and a strong pump pulse are present. The theory is developed for both the velocity and the dipole gauge, which are two complementary ways of incorporating the coupling of the pump pulse to the system described by TB models, especially in a first-principle context. The key link between the two gauges is a unitary transformation in the space of localized WFs, which ensures (approximate) gauge invariance for if the WFs form (approximately) a complete set. 

We applied the theory to subcycle time-resolved photoemission from graphene driven by a strong THz pulse, using a five-band TB model derived from first principles. While the absolute intensity differs in both gauges for a strong pump pulse -- an effect that can mostly be traced back to the additional approximations to photoemission matrix elements -- the photodressed bands agree well. In future work we are planning to compare to first-principle pump-probe spectra within the TDDFT framework~\cite{de_giovannini_first-principles_2017} to benchmark the velocity and the dipole gauge treatment. 

We also discussed typical effects apparent in trARPES in the subcycle regime. The time-dependent pump vector potential photodresses the band structure, which can be understood in the adiabatic picture. Additional features are dynamical broadening and a side band structure which can both be attributed to the pump field varying during the probe pulse. 

Our theory is a an excellent starting point for including realistic electron-electron and electron-phonon scattering within the td-NEGF framework, which will allow for a direct comparison with recent and ongoing experiments on THz dynamics in graphene and topological systems.

\section*{Acknowledgments}
We thank the Stanford Research Computing Center for providing computational resources. Data used in this manuscript is stored on Stanford's Sherlock computing cluster. Supported by the U.S. Department of Energy (DOE), Office of Basic Energy Sciences, Division of Materials Sciences and Engineering, under contract DE-AC02-76SF00515.
M.~S.~thanks the Alexander von Humboldt
Foundation for its support with a Feodor Lynen scholarship.
M.~A.~S.~acknowledges financial support through the Deutsche Forschungsgemeinschaft (DFG, German Research Foundation) via the Emmy Noether program (SE 2558/2).

\appendix

\section{Time-dependent gauge transformation of the Green's function\label{app:gf_trans_gen}}

In this appendix we show how the time-dependent GFs in the velocity gauge (determined by the KBE~\eqref{eq:contour_kb}) and the dipole gauge are related. As the unitary transformation connecting the single-particle Hamiltonian~\eqref{eq:sp_ham_velo} and \eqref{eq:ham_dip} can only be defined in the basis of localized WFs, let us introduce the Wannier GF
\begin{align}
	\label{eq:gf_wan}
	G_{m \vec{R} m^\prime \vec{R}^\prime}(z,z^\prime) = -i \langle T_{\mathcal{C}} \hc_{m \vec{R}}(z) \hcd_{m^\prime \vec{R}^\prime}(z^\prime) \rangle \ ,
\end{align}
where $\hc_{m \vec{R}}$ stands for the annihilation operator with respect to the WFs $|m \vec{R}\rangle$. The time evolution is assumed to be determined by the Hamiltonian~\eqref{eq:hammb_velo}. Translational invariance implies $G_{m \vec{R} m^\prime \vec{R}^\prime}(z,z^\prime) = G_{m 0 m^\prime \vec{R}^\prime-\vec{R}}(z,z^\prime)$, which connects to the band-space GF~\eqref{eq:gf_velo} by 
\begin{align}
	G_{\alpha\alpha^\prime}(\vec{k}; z, z^\prime) = \sum_{\vec{R}}e^{i \vec{k}\cdot\vec{R}}\sum_m C^*_{m \alpha}(\vec{k}) G_{m 0 m^\prime \vec{R}}(z,z^\prime)C_{m^\prime \alpha^\prime}(\vec{k}) \ .
\end{align}
Similary, we can define the dipole-gauge Wannier GF $\widetilde{G}_{m \vec{R} m^\prime \vec{R}^\prime}(z,z^\prime)$, which is defined as Eq.~\eqref{eq:gf_wan}, but with the Hamiltonian~\eqref{eq:hammb_dip} determining the time evolution of the fermionic operators. The operators in the dipole gauge are related to the velocity gauge by the unitary transformation $\hat{U}(t)=e^{\hat{S}(t)}$ with the generator~\eqref{eq:sgen}:
\begin{align}
	\hc_{m \vec{R}}(z)\big|_\mathrm{DG} &=  \hat{U}(t) \hc_{m \vec{R}}(z)\hat{U}^\dagger(z)\big|_\mathrm{VG} \nonumber \\ &= \sum_{n_1 \vec{R}_1} U^*_{n_1 \vec{R}_1 m \vec{R}}(z) \hc_{n_1 \vec{R}_1}(z)\big|_\mathrm{VG}  \ .
\end{align}
Hence, the dipole-gauge GF in the Wannier basis transforms as 
\begin{align}
	\label{eq:gf_gaugetr}
	\widetilde{G}_{m \vec{R} m^\prime \vec{R}^\prime}(z,z^\prime) &= \sum_{\vec{R}_1 \vec{R}_2}\sum_{n_1 n_2} U^*_{n_1 \vec{R}_1 m \vec{R}}(z)
	G_{n_1 \vec{R}_1 n_2 \vec{R}_2}(z,z^\prime) \nonumber \\ &\quad\times U_{n_2 \vec{R}_2 m^\prime \vec{R}^\prime}(z^\prime) \ .
\end{align}

\section{Momentum conservation in the dipole gauge\label{app:mom_dip}}

Inspecting the photoemission matrix elements~\eqref{eq:tdmel} it seems that the momenta are shifted $\vec{k} \rightarrow \vec{k} - q \vec{A}(t)$. This does not result in a shift of the trARPES spectra by $-q\vec{A}(t)$ -- which would be inconsistent with the velocity gauge --  as shown below. Inserting the Wannier representation for $|\phi_{\vec{k}m}\rangle$, we find
\begin{widetext}
\begin{align}
	\label{eq:tdmel_detail}
	\widetilde{M}_m(\vec{k},\vec{p},t) &= \frac{1}{\sqrt{N}}\sum_{\vec{R}}e^{i (\vec{k} - q \vec{A}(t))\cdot \vec{R}}\int d \vec{r}\, \chi^*_{\vec{p}}(\vec{r}) \vec{e}\cdot\hat{\vec{p}} e^{i q \vec{A}(t)\cdot\vec{r}} \phi_m(\vec{r}-\vec{R}) = \frac{1}{\sqrt{N}}\sum_{\vec{R}} 
	e^{i \vec{k}\cdot \vec{R}} \int d \vec{r}\, \chi^*_{\vec{p}}(\vec{r}) \vec{e}\cdot\hat{\vec{p}} e^{i q \vec{A}(t)\cdot(\vec{r}-\vec{R})} 
	\phi_m(\vec{r}-\vec{R}) \nonumber \\ &= \frac{1}{\sqrt{N}}\sum_{\vec{R}} 
	e^{i \vec{k}\cdot \vec{R}} \int d \vec{r}\, \chi^*_{\vec{p}}(\vec{r}+\vec{R}) \vec{e}\cdot\hat{\vec{p}} e^{i q \vec{A}(t)\cdot\vec{r}} \phi_m(\vec{r}) \ ,
\end{align}
\end{widetext}
where $\phi_m(\vec{r}) = \langle \vec{r} | m \vec{R}\rangle$. Exploiting the in-plane Bloch periodicity of the photoelectron states $\chi_{\vec{p}}(\vec{r} + \vec{R}) = e^{i \vec{p}_\parallel \cdot \vec{R}} \chi_{\vec{p}}(\vec{r})$, the integral in Eq.~\eqref{eq:tdmel_detail} turns out to be independent of $\vec{R}$. The sum over $\vec{R}$ reduces to $\sum_{\vec{R}} e^{i (\vec{k} - \vec{p}_\parallel)\cdot \vec{R}} = N \delta_{\vec{k}+\vec{G},\vec{p}_\parallel}$. Therefore, the dipole-gauge matrix elements obey the same momentum conservation as the regular time-independent photoemission matrix elements~\eqref{eq:pes_mel}.

\section{Gauge-invariance of the photocurrent\label{app:trarpes_gauge}}

For showing that the expression~\eqref{eq:trarpes_velo} and \eqref{eq:trarpes_dip} are identical, we consider the integrand kernels
\begin{align}
	\label{eq:kern_velo}
	J_{\vec{p}}(t,t^\prime) = \sum_{\vec{k}}\sum_{\alpha \alpha^\prime} M^*_{\alpha}(\vec{k},\vec{p})
	M_{\alpha^\prime}(\vec{k},\vec{p}) G^<_{\alpha^\prime\alpha}(\vec{k}; t^\prime, t) 
\end{align}
and 
\begin{align}
	\label{eq:kern_dip}
	\widetilde{J}_{\vec{p}}(t,t^\prime) = \sum_{\vec{k}}\sum_{m m^\prime} \widetilde{M}^*_{m}(\vec{k},\vec{p},t)\widetilde{M}_{m^\prime}(\vec{k},\vec{p},t^\prime) \widetilde{G}^<_{m^\prime m}(\vec{k}; t^\prime, t) \ .
\end{align}
It is sufficient to show $\widetilde{J}_{\vec{p}}(t,t^\prime) = J_{\vec{p}}(t,t^\prime)$. Inserting the time-dependent matrix elements~\eqref{eq:tdmel} (for brevity we define $\hat{\Delta} = \vec{e}\cdot \hat{\vec{p}}$) and the Wannier representation one obtains
\begin{widetext}
\begin{align*}
	\widetilde{J}_{\vec{p}}(t,t^\prime) &= \sum_{\vec{k}}\sum_{m m^\prime} \langle \chi_{\vec{p}}| \hat{\Delta}e^{i q \vec{A}(t^\prime)\cdot\vec{r}}| \phi_{\vec{k} - q\vec{A}(t^\prime)m^\prime} \rangle \widetilde{G}^<_{m^\prime m}(\vec{k}; t^\prime, t) \langle \phi_{\vec{k} - q\vec{A}(t)m} | e^{- i q \vec{A}(t)\cdot\vec{r}} \hat{\Delta}^\dagger| \chi_{\vec{p}} \rangle \\
	&= \frac{1}{N}\sum_{\vec{R}\vec{R}^\prime} \sum_{m m^\prime} \langle \chi_{\vec{p}}| \hat{\Delta}e^{i q \vec{A}(t^\prime)\cdot(\vec{r} - \vec{R}^\prime)} | m^\prime \vec{R}^\prime\rangle \widetilde{G}^<_{m^\prime m}(\vec{k}; t^\prime, t) \langle m \vec{R}| e^{-i q\vec{A}(t)\cdot(\vec{r} - \vec{R})} \hat{\Delta}^\dagger | \chi_{\vec{p}} \rangle e^{i\vec{k}\cdot(\vec{R}^\prime - \vec{R})} \ .
\end{align*}
We recognize that the GF in Wannier basis appears explicitly: $\widetilde{G}^<_{m^\prime \vec{R}^\prime m \vec{R}}(t^\prime, t) = (1/N)\sum_{\vec{k}} e^{i\vec{k}\cdot(\vec{R}^\prime - \vec{R})} \widetilde{G}^<_{m^\prime m}(\vec{k}; t^\prime, t)$. Inserting in the above expression yields
\begin{align*}
	\widetilde{J}_{\vec{p}}(t,t^\prime) &= \sum_{\vec{R}\vec{R}^\prime} \sum_{m m^\prime} \langle \chi_{\vec{p}}| \hat{\Delta}e^{i q \vec{A}(t^\prime)\cdot(\vec{r} - \vec{R}^\prime)} | m^\prime \vec{R}^\prime\rangle \widetilde{G}^<_{m^\prime \vec{R}^\prime m \vec{R}}(t^\prime, t) \langle m \vec{R}| e^{-i q\vec{A}(t)\cdot(\vec{r} - \vec{R})} \hat{\Delta}^\dagger | \chi_{\vec{p}} \rangle \\
	&= \sum_{\vec{R}\vec{R}^\prime} \sum_{m m^\prime} \sum_{\vec{R}_1 \vec{R}_2}\sum_{n_1 n_2} \langle \chi_{\vec{p}}| \hat{\Delta} | n_1 \vec{R}_1\rangle U^*_{m^\prime \vec{R}^\prime n_1 \vec{R}_1}(t^\prime) \widetilde{G}^<_{n_1\vec{R}_1 n_2 \vec{R}_2}(t^\prime, t) U_{m \vec{R} n_2\vec{R}_2}(t) \langle n_2 \vec{R}_2 | \hat{\Delta}^\dagger | \chi_{\vec{p}} \rangle \ ,
\end{align*}
\end{widetext}
where $U_{m \vec{R} m^\prime \vec{R}^\prime}(t)$ is defined in Sec.~\ref{subsec:dipgauge}. Substitutung Eq.~\eqref{eq:gf_gaugetr} we find
\begin{align*}
	\widetilde{J}_{\vec{p}}(t,t^\prime) = \sum_{\vec{R}_1 \vec{R}_2}\sum_{n_1 n_2} \langle \chi_{\vec{p}}| \hat{\Delta} | n_1 \vec{R}_1\rangle G^<_{n_1 \vec{R}_1 n_2 \vec{R}_2}(t^\prime, t) \langle n_2 \vec{R}_2 | \hat{\Delta}^\dagger | \chi_{\vec{p}} \rangle \ .
\end{align*}
Using the transformation from Wannier to momentum space (inverse transformation of Eq.~\eqref{eq:gf_wan}), we finally arrive at $\widetilde{J}_{\vec{p}}(t,t^\prime) = J_{\vec{p}}(t,t^\prime)$, thus prooving the gauge invariance.


%

\end{document}